# Non-Diophantine arithmetic as the mathematical foundation for quantum field theory

Mark Burgin and Felix Lev

UCLA, Los Angeles, CA 90095, USA

**Abstract**

The problem of infinities in quantum field theory (QFT) is a long standing problem in particle physics. For solving this problem, different renormalization techniques have been suggested but the problem still exists. Here we suggest another approach to elimination of infinities in QFT, which is based on non-Diophantine arithmetics – a novel mathematical area that already found useful applications in physics. To achieve this goal, new non-Diophantine arithmetics are constructed and their properties are studied. This allows using these arithmetics for computing integrals describing Feynman diagrams. Although in the conventional QFT these integrals diverge, their non-Diophantine counterparts are convergent and rigorously defined.

## 1. Introduction

The phenomenon of quantum field theory (QFT) has no analogs in the history of science. There is no branch of science where so impressive agreements between theory and experiment have been achieved. At the same time, the level of mathematical rigor in QFT is very poor and, as a result, QFT has several known difficulties and inconsistencies.

One of the key mathematical problems of QFT is the problem of infinities: the theory gives divergent expressions for the S-matrix in perturbation theory. In the renormalized theories, the divergencies can be eliminated by using certain rules for operation with divergent integrals. Although those rules are not well substantiated mathematically, in some cases they result in excellent agreement with experiment. Probably, the most famous case is that the results for the electron and muon magnetic moments obtained in quantum electrodynamics (QED) at the end of the 40th agree with experiment at least with the accuracy of eight decimal digits (see, however, a discussion in [1]). In view of this and other successes of QFT, most physicists believe that agreement with experiment is much more important than the rigorous mathematical substantiation.

At the same time, in the non-renormalizable theories, infinities cannot be eliminated by the renormalization procedure, and this is a great obstacle for constructing quantum gravity. As the famous physicist and the Nobel Prize laureate Steven Weinberg wrote in his book [2]: *"Disappointingly this problem appeared with even greater severity in the early days of quantum theory, and although greatly ameliorated by subsequent improvements in the theory, it remains with us to the present day"*. Weinberg's paper "Living with infinities" also expresses this opinion [3].

The main reason of appearance of infinities in QFT is as follows. QFT works with quantized fields $\varphi(x)$ (where $x$ is treated as a point in Minkowski space), while, for interacting fields, the Lagrangian contains products of fields at the same points. However, as pointed out, e.g., in the textbook [4], interacting quantized fields necessarily should be treated as operator distributions, and the known fact of the theory of distributions is that their products at the same points are not well defined [18, 19].

Physicists usually ignore this fact and believe that such products are needed for preserving locality. However, $\varphi(x)$ is an operator in the Fock space for an infinite number of particles. Each particle in the Fock space can be described by its own coordinates (in the approximation when the position operator exists, cf., e.g., [5]). In view of this fact, the following natural question arises: why do we need an extra coordinate $x$ which does not belong to any particle? This coordinate does not have a physical meaning and is simply a parameter arising from the quantization of the classical field.

So, even the above example explains why QFT is not based on rigorous mathematics and why, nevertheless, physicists adopt this construction and try to exclude infinities only *after* this (mathematically incorrect) construction has been performed.

The problem of excluding infinities in QFT has been discussed in a vast literature. Most authors tried to solve this problem in the framework of standard mathematics involving infinitesimals but, to the best of our knowledge, the problem has not been solved in such a way. On the other hand, several authors proposed versions of mathematics where infinitesimals do not exist in principle. One of the approaches is mathematics based on non-Diophantine arithmetics (see e.g., [6] and references therein), and another one is based on applications of finite mathematics to physics (see e.g., [7] and references therein). Those approaches are considerably different, and at the present stage of particle theory, it is not clear whether there exists an approach which can resolve all difficulties of standard quantum theory. That is why different approaches should be investigated. In the present paper we show that the approach based on non-Diophantine arithmetics can resolve inconsistencies arising in perturbation theory of QFT.

The principal difference between the approach suggested in [7] and conventional ones can be explained as follows. Consider some physical quantity, e.g., the magnitude of the four-dimensional momentum $p$. In each experiment this quantity is always within some finite range, but physicists usually believe that the theory should not involve a quantity $p_{max}$, which is the absolute maximum of all possible momenta in nature, i.e., that in *any* phenomenon $p$ cannot exceed $p_{max}$. This agrees with the belief of most physicists that the conventional Diophantine mathematics involving infinitesimals describes nature adequately. On the other hand, in the approaches [7], any physical quantity cannot exceed some finite absolute maximum of this quantity in all possible phenomena.

A problem arises whether one can explicitly specify the values of the absolute maximum for each physical quantity. For example, in the approach suggested in [7], quantum theory is based on a finite ring, and no dimensionless physical quantity can exceed the characteristic of this ring. It is argued that the characteristic depends on the current state of the universe, i.e., the characteristic is different in different stages of the universe. It has been shown that the quantity *G*, called the gravitational constant, is not a constant but is a function of the characteristic. Then, comparing the expression for *G* obtained in [7] with the experimental value, one can see that, at the present stage of the universe, the characteristic of the utilized ring is a huge number of the order of *exp*($10^{80}$) or more. One might think that, since this number is so huge, one can treat it as infinitely large. However, since the gravitational constant *G* depends on the natural logarithm of the characteristic, which is "only" of the order of $10^{80}$ or more, gravity is observable.

In the present paper, we describe how it is possible to evade infinities in physical theories using non-Diophantine arithmetics. In this context, taking into account ideas of Kronecker about building the whole mathematics based on arithmetic, we call mathematics based on non-Diophantine arithmetics by the name non-Diophantine mathematics or NDM, which provides various opportunities for the further development of physics [6].

In view of the above discussion, the present paper is organized as follows. In Section 2, we consider typical divergent integrals which appear in QFT as a consequence of the fact that standard approach to QFT is not well defined. In Section 3, special non-Diophantine arithmetics are constructed with the aim of divergence elimination in quantum theories. In Section 4, we explain that based on the constructed non-Diophantine arithmetics, divergent integrals from QFT can be considered without divergence in the framework of a consistent mathematical theory. Finally, in Section 5, we discuss further possibilities to achieve higher mathematical rigor in QFT.

## 2. Divergent integrals in QED

In QED, a typical divergent integral arises as follows (see e.g., standard textbooks [8, 9] and others). In the second order of perturbation theory, the photon propagator is defined by the polarization operator

$$\Pi(k)^{\alpha\beta} = i \int Tr\{\gamma^\alpha (\hat{p} + \hat{k} - m_0 + i\varepsilon)^{-1} \gamma^\beta (\hat{p} - m_0 + i\varepsilon)^{-1}\} d^4p/(2\pi)^4 \quad (1)$$

Here $p$ and $k$ are the four-momenta, $p = (p_0, p_1, p_2, p_3)$ and analogously for $k$. The indices $\alpha$, $\beta$ take the values (0, 1, 2, 3), $\gamma$ is the Dirac gamma matrix, $d^4p$ is the volume element in the momentum space, $Tr$ is the trace of the matrix, and the Minkowski metric tensor is diagonal such that $\eta_{00} = -\eta_{11} = -\eta_{22} = -\eta_{33} = 1$, $\hat{p} = \gamma^\alpha p_\alpha$ where the sum over repeated indices is assumed. The mass of the bare electron $m_0$ is taken with a small imaginary correction $-i\varepsilon$ ($\varepsilon > 0$) for avoiding singularities resulting from zeros of the denominators in formula (1), and in the final result, the limit $\varepsilon \to 0$ should be taken.

By using standard expression for traces of gamma matrices, formula (1) can be represented as

$$\Pi(k)^{\alpha\beta} = i\int [2p^\alpha p^\beta + (m_0^2 - p^2 + k^2/4)\eta^{\alpha\beta} - k^\alpha k^\beta/4][(p+k/2)^2 - m_0^2 + i\varepsilon]^{-1}[(p-k/2)^2 - m_0^2 + i\varepsilon]^{-1} d^4p/(2\pi)^4 \quad (2)$$

By expanding the integrand in powers of $k$, it becomes clear that the most divergent part of $\Pi(k)^{\alpha\beta}$ is

$$\Pi^{\alpha\beta} = i\int [2p^\alpha p^\beta - p^2\eta^{\alpha\beta}][p^2 - m_0^2 + i\varepsilon]^{-2} d^4p/(2\pi)^4 = -(i/2)\Pi\eta^{\alpha\beta} \quad (3)$$

and

$$\Pi = i\int [p^2 - m_0^2 + i\varepsilon]^{-2} p^2 d^4p/(2\pi)^4 \quad (4)$$

Here the integral for $\Pi$ quadratically diverges at large momenta $p$. Note that in the Minkowski metric, $p^2 = p_0^2 - \mathbf{p}^2$ where $\mathbf{p}$ is the vector $(p^1, p^2, p^3)$ in the three-dimensional Euclidean space.

In the literature, the expression for $\Pi$ is treated in different ways. For example:

● One can insert into the integrand the factor $\Lambda^4/(\Lambda^2-p^2)^2$ where $\Lambda$ has a large value. Then $\Pi$ becomes convergent but its limit at $\Lambda \to \infty$ is infinite.

● One can represent each multiplier $(p^2-m_0^2+i\varepsilon)^{-1}$ in the denominator of formula (2) as

$$(p^2-m_0^2+i\varepsilon)^{-1} = -i\int_0^\infty exp[i(p^2-m_0^2+i\varepsilon)x]\mathrm{d}x,$$

change the order of integration and then the integral over $p$ becomes Gaussian. However, the result will become infinite in the limit $\varepsilon \to 0$.

● If one defines $p_4 = ip_0$, then the four-momentum $p$ becomes the vector $(p_0, \mathbf{p})$ in the Euclidean space such that $p^2 = p_4^2 + \mathbf{p}^2$. Then the correction $i\varepsilon$ is not needed, and we have

$$\Pi = i\int (p^2+m_0^2)^{-2} p^2 \mathrm{d}^4p/(2\pi)^4 \qquad (5)$$

This integral also quadratically diverges. It can be calculated if the integration over $|p|$ from zero to infinity is replaced by integration from zero to some parameter $p_{max}$ (see below), but the final result will become infinite at $p_{max} \to \infty$.

All the above ways of treating infinities are not mathematically correct because from the very beginning they involve expressions which are not well defined.

In what follows, we will consider four-momenta in the Euclidean metric. Then, as shown in standard textbooks on QED, in the second order of perturbation theory the following divergent integrals appear (see e.g., Chapter 36 in [8]):

$$I_1 = \int (p^2+l)^{-2} d^4p; \quad I_2 = \int (p^2+l)^{-2} p^2 d^4p; \quad I_3 = \int (p^2+l)^{-3} p^2 d^4p \qquad (6)$$

Here $I_1$ is the divergent contribution to the Feynman diagram describing the electron self energy, $I_2$ is the divergent contribution to the Feynman diagram describing the photon self energy, and $I_3$ is the divergent contribution to the Feynman diagram describing the electron-photon vertex. In section 4 we describe how these integrals are treated in both, standard mathematics and non-Diophantine mathematics.

## 3. Non-Diophantine arithmetics for finite physics

As our goal is to create quantum field theory without divergence leading to the actual infinity, we construct non-Diophantine arithmetics with upper and lower boundaries.

A regular way of constructing non-Diophantine arithmetics is to take a set of ordinary numbers, e.g., natural, integer or real numbers, and to establish its connections to the conventional Diophantine arithmetic in such a way that performance of operations in the non-Diophantine arithmetic would be induced by corresponding operations in the Diophantine arithmetic.

At first, we build a special non-Diophantine arithmetic $A$ of integer numbers using weak projectivity with the Diophantine arithmetic $Z$ of all integer numbers [6, 11, 12]. To do this, we remind the definition of weak projectivity.

Let us take two abstract arithmetics $A_1 = (A_1; +_1, \circ_1, \leq_1)$ and $A_2 = (A_2; +_2, \circ_2, \leq_2)$ and consider two mappings $g: A_1 \to A_2$ and $h: A_2 \to A_1$.

**Definition 3.1.** a) An abstract arithmetic $A_1 = (A_1; +_1, \circ_1, \leq_1)$ is called *weakly projective* with respect to an abstract arithmetic $A_2 = (A_2; +_2, \circ_2, \leq_2)$ if there are following relations between orders and operations in $A_1$ and in $A_2$:

$$a +_1 b = h(g(a) +_2 g(b))$$

$$a \circ_1 b = h(g(a) \circ_2 g(b))$$

$$a \leq_1 b \text{ only if } g(a) \leq_2 g(b)$$

b) The mapping $g$ is called the *projector* and the mapping $h$ is called the *coprojector* for the pair $(A_1, A_2)$.

The functions $g$ and $h$ determine a *weak projectivity* between the arithmetic $A_1$ and the arithmetic $A_2$.

Informally, it means that to perform an operation, e.g., addition or multiplication, in $A_1$ with two numbers $a$ and $b$, we map these numbers into $A_2$, perform this operation there, and map the result back to $A_1$.

For instance, let us take $A_2 = Z$, $g(x) = x + 1$ and $h(x) = x - 1$. Taking $a = 2$ and $b = 3$, we have

$$2 +_1 3 = h(g(a) + g(b)) = h(g(2) + g(3)) = h(3 + 4) = h(7) = 6$$

In such a way, these two functions $g$ and $h$ define the non-Diophantine arithmetic $A_1$ of integer numbers. Note that contains the same integer numbers as the conventional arithmetic $Z$ but operations with them are defined in a different way.

Let us take a natural number $L$ as the boundary parameter of the non-Diophantine arithmetic $A_L$. We build the non-Diophantine arithmetic $A_L$ taking the following functions $g$ and $h$ for establishing a weak projectivity between $A_L$ and $Z$.

$$g(x) = \begin{cases} x & \text{when } -L \leq x \leq L \\ L & \text{when } x > L \\ -L & \text{when } x < -L \end{cases}$$

and

$$h(x) = \begin{cases} x & \text{when } -L \leq x \leq L \\ L & \text{when } x > L \\ -L & \text{when } x < -L \end{cases}$$

Then the operations, that is, addition, subtraction, and multiplication, in $A_L$ are defined in the following way:

$$a +_L b = h(g(a) + g(b))$$

$$a \times_L b = h(g(a) \times g(b))$$

$$a -_L b = h(g(a) - g(b))$$

We call the number $L$ by the name the *upper boundary number* of the arithmetic $A_L$.

Note that the non-Diophantine arithmetic $A_L$ contains all integer numbers but if we are inside the interval $(-L, L)$, then only numbers larger than $-L$ and smaller than $L$ are accessible. All other integer numbers do not impact operations with accessible numbers. As a result such arithmetics $A_L$ exactly model computer arithmetics with integer numbers [17, 18]. It is also possible to suggest that such arithmetics $A_L$ will be useful for building finite physics based on sound and adequate mathematical structures.

Contemporary physical theories suggest that it might be reasonable to take $L = 10^{100}$. Let us consider examples of operations in this arithmetic where $\oplus$ denotes addition, $\ominus$ denotes subtraction, and $\otimes$ denotes multiplication in the arithmetic $A_{10^{100}}$.

$$1000 \oplus 1000 = 2000$$
$$10^{90} \oplus 10^{90} = 2 \times 10^{90}$$
$$10^{200} \oplus 10^{10} = 10^{100}$$
$$1000 \otimes 1000 = 1000000$$
$$10^{90} \otimes 10^{90} = 10^{100}$$
$$10^{200} \otimes 10^{10} = 10^{100}$$
$$10^{90} \ominus 10^{80} = (10^{10} - 1)10^{80}$$
$$10^{200} \ominus 10^{10} = 10^{100}$$
$$(10^{200} + 1000) \ominus 10^{200} = 10^{100}$$

Let us study properties of the arithmetic $A_L$. Direct application of the definition of operations in the arithmetic $A_L$ gives us the following result.

**Proposition 3.1.** For any natural numbers $L$ and $n$, we have the following identities in the arithmetic $A_L$:

$$L +_L n = L$$

$$L \times_L n = L$$

$$-L +_L (-n) = -L -_L n = -L$$

$$-L \times_L n = -L$$

$$-n +_L n = n -_L n = 0$$

$$L \times_L (-n) = -L$$

$$n -_L L = 0 \text{ if } n > L$$

$$0 \times_L n = 0$$

$$0 +_L n = n \text{ if } -L \leq n \leq L$$

$$0 +_L n = L \text{ if } n > L$$

$$0 +_L (-n) = -L \text{ if } n > L$$

**Theorem 3.1.** For any natural number $L$, we have:
a) addition and multiplication are commutative in the arithmetic $A_L$;
b) addition in the arithmetic $A_L$ is not always associative;
c) multiplication in the arithmetic $A_L$ is always associative;
d) multiplication in the arithmetic $A_L$ is not always distributive with respect to addition;
e) The results of addition, subtraction, and multiplication in the arithmetic $A_L$ cannot be larger than $L$ and less than $-L$.

*Proof.* a) Proposition 2.2.4 [6] implies that addition is commutative in the arithmetic $A_L$. Proposition 2.2.5 from [6] implies that multiplication is commutative in the arithmetic $A_L$;

b) Let us take the arithmetic $A_{10}$ with addition $\oplus$ and consider the following expressions.

$$(5 \oplus 7) \oplus (-8) = 10 \oplus (-8) = 2$$

At the same time, we have

$$5 \oplus (7 \oplus (-8)) = 5 \oplus (-1) = 4$$

So, addition is not associative in the arithmetic $A_{10}$.

c) Let us consider the associative identity where *a*, *b* and *c* are natural numbers

$$(a \times_L b) \times_L c = a \times_L (b \times_L c)$$

If one or more of these numbers is larger than or equal to *L*, then both sides are equal to *L*.

If $a \times_L b$ is equal to *L*, then the left side equal to *L* by Proposition 3.1 and the right side equal to *L* because $a \times_L b$ is less than or equal to $a \times_L (b \times_L c)$.

In all other cases, we have the conventional multiplication of natural numbers, which is associative.

When some of the numbers *a*, *b* and *c* are negative, the inference is the same because multiplication is performed separately with signs and separately with numbers.

d) Let us take the arithmetic $A_{10}$ with addition $\oplus$ and multiplication $\otimes$ and consider the following expressions.

$$3 \otimes (7 \oplus (-5)) = 3 \otimes 2 = 6$$

At the same time, we have

$$(3 \otimes 7) \oplus (3 \otimes (-5)) = 10 \oplus (-10) = 0$$

So, multiplication is not distributive with respect to addition in the arithmetic $A_{10}$.

e) This statement follows from definitions because the function *h* is bounded by *L* from above and by -*L* from below.

Theorem is proved.

Also note that the Diophantine arithmetic **Z** of all integer numbers is, in some sense, the limit of arithmetics $A_L$ when *L* tends to infinity.

Let us study properties of the subarithmetic $A_L^+$ of all positive numbers from the arithmetic $A_L$.

**Theorem 3.2.** For any natural number *L*, we have:
a) addition and multiplication are commutative in the arithmetic $A_L^+$;
b) addition and multiplication are associative in the arithmetic $A_L^+$;
c) multiplication in the arithmetic $A_L^+$ is distributive with respect to addition;
d) addition and multiplication are monotone operations in the arithmetic $A_L^+$.

*Proof.* a) Addition and multiplication are commutative in the arithmetic $A_L^+$ because subarithmetics inherit commutativity of from their super-arithmetic [6] and by definition, $A_L^+$ is a subarithmetic of $A_L$ where by Theorem 3.1, addition and multiplication are commutative.

b) Let us consider the associative identity for addition in $A_L^+$ where *a*, *b* and *c* are natural numbers

$$(a +_L b) +_L c = a +_L (b +_L c)$$

If one or more of these numbers is larger than or equal to *L*, then by Proposition 3.1, both sides are equal to *L*.

If $a +_L b$ is equal to *L*, then the left side equal to *L* by Proposition 3.1 and the right side equal to *L* because $a +_L b$ is less than or equal to $a +_L (b +_L c)$.

In all other cases, we have the conventional addition of natural numbers, which is associative.

Let us consider the associative identity for multiplication in $A_L^+$ where *a*, *b* and *c* are natural numbers

$$(a \times_L b) \times_L c = a \times_L (b \times_L c)$$

If one or more of these numbers is larger than or equal to *L*, then both sides are equal to *L*.

If $a \times_L b$ is equal to *L*, then the left side equal to *L* by Proposition 3.1 and the right side equal to *L* because $a \times_L b$ is less than or equal to $a \times_L (b \times_L c)$.

In all other cases, we have the conventional multiplication of natural numbers, which is associative.

c) Let us consider the distributive identity for multiplication in $A_L^+$ where *a*, *b* and *c* are natural numbers

$$a \times_L (b +_L c) = (a \times_L b) +_L (a \times_L c)$$

If one or more of these numbers is larger than or equal to *L*, then both sides are equal to *L*.

If $b +_L c$ is equal to *L*, then the left side equal to *L* by Proposition 3.1 and the right side equal to *L* because $a \times_L b$ is larger than or equal to *b* and $a \times_L c$ is larger than or equal to *c*, which implies that $(a \times_L b) +_L (a \times_L c)$ is larger than or equal to $b +_L c$, which in turn is equal to *L*,

In all other cases, we have the conventional operations with natural numbers, in which multiplication is distributive with respect to addition.

d) At first, we consider addition in $A_L^+$. Let us assume that $c \leq b$ and compare $a +_L b$ and $a +_L c$. Note that by Proposition 3.1, the product or sum of two natural numbers from $A_L^+$ cannot be larger than $L$.

If $L \leq a$, then $a +_L b = L$ and $a +_L c = L$. Consequently, $a +_L c \leq a +_L b$.

If $L \leq b$, then $a +_L b = L$ and $a +_L c \leq L = a +_L b$.

If $L \leq c$, then $L \leq b$, $a +_L b = L$ and $a +_L c = L \leq L = a +_L b$.

If $a, b, c < L$ and $a +_L b = L$, then $a +_L c \leq L = a +_L b$.

If $a, b, c < L$ and $a +_L c = L$, then $b = c +_L k = c + k$ and by associativity of addition (part b of this theorem), we have

$$a +_L b = a +_L (c +_L k) = (a \times_L c) +_L k = L +_L k = L$$

Consequently, $a +_L c \leq a +_L b$.

If $a, b, c < L$, $a +_L b = L$, and $a +_L c < L$, then all operations are performed in the Diophantine arithmetic of natural numbers where addition is a monotone operation. Consequently, we have $a +_L c \leq a +_L b$.

Thus, monotonicity of addition in the arithmetic $A_L^+$ is proved.

Now we treat multiplication in $A_L^+$. Let us assume that $c \leq b$ and compare $a \times_L b$ and $a \times_L c$. Note that by Proposition 3.1, the product or sum of two natural numbers from $A_L^+$ cannot be larger than $L$.

If $L \leq a$, then $a \times_L b = L$ and $a \times_L c = L$. Consequently, $a \times_L c \leq a \times_L b$.

If $L \leq b$, then $a \times_L b = L$ and $a \times_L c \leq L = a \times_L b$.

If $L \leq c$, then $L \leq b$, $a \times_L b = L$ and $a \times_L c = L \leq L = a \times_L b$.

If $a, b, c < L$ and $a \times_L b = L$, then $a \times_L c \leq L = a \times_L b$.

If $a, b, c < L$ and $a \times_L c = L$, then $b = c +_L k = c + k$ and by distributivity, we have

$$a \times_L b = a \times_L (c + k) = (a \times_L c) +_L (a \times_L k) = L +_L (a \times_L k) = L$$

Consequently, $a \times_L c = L \leq a \times_L b = L$.

If $a$, $b$, $c < L$, $a \times_L b = L$, and $a \times_L c < L$, then all operations are performed in the Diophantine arithmetic of natural numbers where multiplication is a monotone operation. Consequently, $a \times_L c \leq a \times_L b$.

Thus, monotonicity of multiplication in the arithmetic $A_L^+$ is proved.

Theorem is proved.

**Remark 3.1**. Although addition and multiplication are monotone operations in the arithmetic $A_L^+$, they are not strictly monotone. Indeed, for any natural number $n$ less than $L$, we have $L + n = L$.

If we want to have more freedom for operating with numbers, we use the identity function $e(x) = x$ instead of the function $g$ for building another non-Diophantine arithmetic $A_{eL}$ by establishing the weak projectivity between $A_L$ and $Z$. The function $h$ remains the same.

$$h(x) = \begin{cases} x & \text{when } -L \leq x \leq L \\ L & \text{when } x > L \\ -L & \text{when } x < -L \end{cases}$$

Then the operations, that is, addition, subtraction, and multiplication, in $A_{eL}$ are defined in the following way:

$$a +_{eL} b = h(e(a) + e(b)) = h(a + b)$$

$$a \times_{eL} b = h(e(a) \times e(b)) = h(a \times b)$$

$$a -_{eL} b = h(e(a) - e(b)) = h(a - b)$$

We call the number $L$ by the name the *upper boundary number* of the arithmetic $A_{eL}$.

Let us consider examples of operations in this arithmetic where $\oplus$ denotes addition, $\ominus$ denotes subtraction, and $\otimes$ denotes multiplication in the arithmetic $A_{e10^{100}}$.

$$1000 \oplus 1000 = 2000$$

$$10^{90} \oplus 10^{90} = 2 \times 10^{90}$$

$$(10^{200} + 1000) \oplus 10^{10} = 10^{100}$$

$$1000 \otimes 1000 = 1000000$$

$$10^{90} \otimes 10^{90} = 10^{100}$$

$$10^{200} \otimes 10^{10} = 10^{100}$$

$$10^{90} \ominus 10^{80} = (10^{10} - 1)10^{80}$$

$$(10^{200} + 1000) \ominus 10^{200} = 1000$$

Many properties of the arithmetic $A_{eL}$ are similar to the properties of the arithmetic $A_L$.

**Proposition 3.2.** For any natural numbers $L$ and $n$, we have the following identities in the arithmetic $A_{eL}$:

$$L +_{eL} n = L$$

$$L \times_{eL} n = L$$

$$-L +_{eL} (-n) = -L -_{eL} n = -L$$

$$-L \times_{eL} n = -L$$

$$-n +_{eL} n = n -_{eL} n = 0$$

$$L \times_{eL} (-n) = -L$$

$$0 \times_{eL} n = 0$$

$$0 +_{eL} n = n \text{ if } -L \leq n \leq L$$

$$0 +_{eL} n = L \text{ if } n > L$$

$$0 +_{eL} (-n) = -L \text{ if } n > L$$

**Theorem 3.3.** For any natural number $L$, we have:
a) addition and multiplication are commutative in the arithmetic $A_{eL}$;
b) addition in the arithmetic $A_{eL}$ is not always associative;
c) multiplication in the arithmetic $A_{eL}$ is always associative;
d) multiplication in the arithmetic $A_{eL}$ is not always distributive with respect to addition;

e) The results of addition, subtraction, and multiplication in the arithmetic $A_{eL}$ cannot be larger than $L$ and less than $-L$.

*Proof* is similar to the proof of Theorem 3.1.

At the same time, the arithmetic $A_{eL}$ has properties such that the arithmetic $A_L$ does not have.

For instance, if $m > n > L$ in the arithmetic $A_L$, then we have

$$m -_L n = h(g(m) - g(n)) = h(L - L) = h(0) = 0$$

In particular, it means that in the arithmetic $A_{10^{100}}$ considered above, we have

$$(10^{100} + 100) \ominus (10^{100} + 50) = 0$$

However, taking the arithmetic $A_{e10^{100}}$ considered above, we see that

$$(10^{100} + 100) \ominus (10^{100} + 50) = h((10^{100} + 100) - (10^{100} + 50)) =$$
$$h(100 - 50) = h(50) = 50$$

Let us consider the subarithmetic $A_{eL}^+$ of all positive numbers from the arithmetic $A_{eL}$.

**Theorem 3.4.** For any natural number $L$, we have:

a) addition and multiplication are commutative in the arithmetic $A_{eL}^+$;

b) addition and multiplication are associative in the arithmetic $A_{eL}^+$;

c) multiplication in the arithmetic $A_{eL}^+$ is distributive with respect to addition;

d) addition and multiplication are monotone operations in the arithmetic $A_{eL}^+$.

*Proof* is similar to the proof of Theorem 3.2.

**Remark 3.2.** Although addition and multiplication are monotone operations in the arithmetic $A_{eL}^+$, they are not strictly monotone. Indeed, for any natural number $n$ less than $L$, we have $L + n = L$.

**Remark 3.3.** It is possible to use arithmetics $A_L$ and $A_{eL}$ as rigorous representations of computer arithmetics. In this case, the number $L$ will usually be much larger than $10^{100}$.

In arithmetic, it is possible to define not only binary operations but operations of higher arities, which are applied to more than two numbers and are often used in other fields of mathematics and in science. There are also integral arithmetical operations, which can be applied to any finite number of numbers [20, 21]. In physics, such integral arithmetical operations as summation and taking powers are very useful. To be able efficiently employ advantages of non-Diophantine arithmetics for developing physical theories, we describe corresponding operations in non-Diophantine arithmetics. While summation is determined in a unique way in the conventional Diophantine arithmetic, there are several ways to define non-Diophantine summation, which is also called non-Diophantine multiple addition [11]. Here we consider two of these operations - sequential summation and parallel summation.

Let us consider a non-Diophantine arithmetic $A$ with addition $\oplus$ and multiplication $\otimes$ and assume that $A$ is weakly projective with respect to the arithmetic $Z$ with the projector $g$ and the coprojector $h$.

**Definition 3.2.** The *sequential sum* is defined in the following way

$$\sum_{i=1}^{n} {}_\oplus a_i = (\ldots((a_1 \oplus a_2) \oplus a_3) \ldots \oplus a_n)$$

**Proposition 3.3.** For any natural number $n$, the sequential sum $\sum_{i=1}^{n} {}_\oplus$ in $A$ is commutative if addition $\oplus$ is associative.

Theorem 3.2 implies the following result.

**Corollary 3.1.** For any natural number $n$, the sequential sum $\sum_{i=1}^{n} {}_\oplus$ is commutative in $A_L{}^+$.

Theorem 3.4 implies the following result.

**Corollary 3.2.** For any natural number $n$, the sequential sum $\sum_{i=1}^{n} {}_\oplus$ is commutative in $A_{eL}{}^+$.

**Definition 3.3.** The *parallel sum* is defined in the following way

$$\sum_{i=1}^{n}{}^{\oplus} a_i = h(g(a_1) + g(a_2) + g(a_3) \ldots + g(a_n))$$

**Proposition 3.4.** For any natural number $n$, the parallel sum $\sum_{i=1}^{n} \oplus$ is always commutative in $A$.

**Corollary 3.3.** For any natural number $n$, the parallel sum $\sum_{i=1}^{n} \oplus$ is always commutative in $A_L$.

**Corollary 3.4.** For any natural number $n$, the parallel sum $\sum_{i=1}^{n} \oplus$ is always commutative in $A_{eL}$.

In the same way as for addition, we define two forms of multiple multiplication - sequential multiple multiplication and parallel multiple multiplication.

**Definition 3.4.** The *sequential product* is defined in the following way.

$$\prod_{i=1}^{n} \oplus a_i = (\ldots ((a_1 \otimes a_2) \otimes a_3) \ldots \otimes a_n)$$

**Proposition 3.5.** For any natural number $n$, the sequential product $\prod_{i=1}^{n} \otimes$ in $A$ is commutative if multiplication $\otimes$ is associative.

Theorem 3.2 implies the following result.

**Corollary 3.5.** For any natural number $n$, the sequential product $\prod_{i=1}^{n} \otimes$ is commutative in $A_L^+$.

Theorem 3.4 implies the following result.

**Corollary 3.6.** For any natural number $n$, the sequential product $\prod_{\otimes i=1}^{n}$ is commutative in $A_{eL}^+$.

In a usual way, powers (exponents) of numbers are defined using multiplication two forms of multiple multiplication - sequential multiple multiplication and parallel multiple multiplication – give us two forms of exponents - sequential exponents and parallel multiple exponents.

**Definition 3.5.** The *sequential power* (*sequential exponent*) of a number $a$ is defined in the following way

$$a^n = \prod_{i=1}^{n} {}_{\oplus} a_i$$

where $a_i = a$ for all $i = 1, 2, 3, \ldots, n$.

**Definition 3.6.** The *parallel product* is defined in the following way

$$\prod_{i=1}^{n} {}^{\oplus} a_i = h(g(a_1) \times g(a_2) \times g(a_3) \times \ldots \times g(a_n))$$

**Proposition 3.6.** For any natural number $n$, the parallel product $\prod_{i=1}^{n} {}^{\otimes}$ in $A$ is always commutative.

**Corollary 3.7.** For any natural number $n$, the parallel product $\prod_{i=1}^{n} {}^{\otimes}$ is always commutative in $A_L$.

**Corollary 3.8.** For any natural number $n$, the parallel product $\prod_{i=1}^{n} {}^{\otimes}$ is always commutative in $A_{eL}$.

**Definition 3.7.** The *parallel power* (*parallel exponent*) of a number $a$ is defined in the following way

$$a^n = \prod_{i=1}^{n} {}_{\oplus} a_i$$

where $a_i = a$ for all $i = 1, 2, 3, \ldots, n$.

Physical theories use real and complex numbers. That is why, we extend the described techniques for building non-Diophantine arithmetics $\mathbf{R}_L$ of real numbers based on weak projectivity with the Diophantine arithmetic $\mathbf{R}$ of all real numbers [6, 10, 11]. To do this, we can use the same functions $g$ and $h$ as in the case of integer numbers. Here $L$, as before, is the boundary parameter of the arithmetic $\mathbf{R}_L$. It can be any positive real number but without diminishing generality, it is possible to assume that $L$ is a natural number.

$$g(x) = \begin{cases} x \text{ when } -L \leq x \leq L \\ L \text{ when } x > L \\ -L \text{ when } x < -L \end{cases}$$

and

$$h(x) = \begin{cases} x \text{ when } -L \leq x \leq L \\ L \text{ when } x > L \\ -L \text{ when } x < -L \end{cases}$$

Then the operations, that is, addition, subtraction, division and multiplication, in $\mathbf{R}_L$ are defined in the following way:

$$a +_L b = h(g(a) + g(b))$$

$$a \times_L b = h(g(a) \times g(b))$$

$$a -_L b = h(g(a) - g(b))$$

$$a \div_L b = h(g(a) \div g(b))$$

Let us consider examples of operations in the arithmetic $\mathbf{R}_{10^{100}}$ where $\oplus$ denotes addition, $\ominus$ denotes subtraction, $\oslash$ denotes division, and $\otimes$ denotes multiplication in the arithmetic $\mathbf{R}_{10^{100}}$.

$$1000 \oplus 1000 = 2000$$
$$10^{90} \oplus 10^{90} = 2 \times 10^{90}$$
$$10^{200} \oplus 10^{10} = 10^{100}$$
$$1000 \otimes 1000 = 1000000$$

$$10^{90} \otimes 10^{90} = 10^{100}$$
$$10^{200} \otimes 10^{10} = 10^{100}$$
$$10^{90} \ominus 10^{80} = (10^{10} - 1)10^{80}$$
$$10^{200} \ominus 10^{10} = 10^{100}$$
$$(10^{200} + 1000) \ominus 10^{200} = 10^{100}$$
$$10^{200} \oslash 10^{10} = 10^{90}$$
$$10^{200} \oslash 10^{100} = 1$$

**Proposition 3.7.** For any natural number $L$, the arithmetic $A_L$ is a subarithmetics of the arithmetic $R_L$.

Although the arithmetic $R_L$ is much larger than the arithmetic $A_L$, they have many similar properties.

**Theorem 3.5.** For any natural number $L$, we have:
  a) addition and multiplication are commutative in the arithmetic $R_L$;
  b) addition in the arithmetic $R_L$ is not always associative;
  c) multiplication in the arithmetic $R_L$ is always associative;
  d) multiplication in the arithmetic $R_L$ is not always distributive with respect to addition.

*Proof* is similar to the proof of Theorem 3.1.

At the same time, the arithmetic $R_L$ has properties such that the arithmetic $A_L$ does not have. For instance, in the arithmetic $A_{el}$, numbers do not inverse numbers while for the arithmetic $R_L$, we have the following result.

**Theorem 3.7.** A number $r$ from the arithmetic $R_L$ has the inverse if and only if $|r| \geq 1/L$.

*Proof. Sufficiency*. Let us take a real number $r$ such that $|r| \geq 1/L$. It is sufficient to consider only positive numbers and in this case, we have two situations: (1) $r \geq L$ and (2) $r < L$.

In the case (1), let us consider multiplication of $r$ by $1/L$. We have

$$r \times_L 1/L = h(g(r) \times g(1/L)) = h(L \times 1/L) = h(1) = 1$$

It means that $1/L$ is the inverse of $r$.

In the case (2), we have

$$r \times_L 1/r = h(g(r) \times g(1/r)) = h(r \times 1/r) = h(1) = 1$$

because $r \geq 1/L$ implies $1/r \leq L$.

Sufficiency is proved.

*Necessity*. Let us take a positive real number $r$ such that $r < 1/L$. Then for any real number $q$ such that $q \leq L$, we have $r \times_L q < 1$.

If a real number $t > L$, then we have

$$r \times_L t = h(g(r) \times g(t)) = h(r \times L) = L$$

Consequently, there are no real number $q$ such that $r \times_L q = 1$.

Theorem is proved.

However, division is defined for all numbers from the arithmetic $\boldsymbol{R}_L$ but 0.

Let us consider the subarithmetic $\boldsymbol{R}_{eL}^+$ of all positive numbers from the arithmetic $\boldsymbol{R}_{eL}$.

**Theorem 3.6.** For any natural number $L$, we have:

a) addition and multiplication are commutative in the arithmetic $\boldsymbol{R}_L$;

b) addition and multiplication are associative in the arithmetic $\boldsymbol{R}_L$;

c) multiplication in the arithmetic $\boldsymbol{R}_L$ is distributive with respect to addition;

d) addition and multiplication are monotone operations in the arithmetic $\boldsymbol{R}_L$.

*Proof* is similar to the proof of Theorem 3.2.

If we want to have more freedom for operating with numbers, we use the identity function $e(x) = x$ instead of $g$ for building another non-Diophantine arithmetic $\boldsymbol{R}_{eL}$ by establishing a weak projectivity between $\boldsymbol{R}_{eL}$ and $\boldsymbol{Q}$. The function $h$ remains the same.

$$h(x) = \begin{cases} x \text{ when } -L \leq x \leq L \\ L \text{ when } x > L \\ -L \text{ when } x < -L \end{cases}$$

Then the operations, that is, addition and multiplication, in $\boldsymbol{R}_{eL}$ are defined in the following way:

$$a +_{eL} b = h(e(a) + e(b)) = h(a + b)$$

$$a \times_{eL} b = h(e(a) \times e(b)) = h(a \times b)$$

$$a -_{eL} b = h(e(a) - e(b)) = h(a - b)$$

$$a \div_{eL} b = h(g(a) \div g(b)) = h(a \div b)$$

Let us consider examples of operations in the arithmetic $\boldsymbol{R}_{e10^{100}}$ where $\oplus$ denotes addition, $\ominus$ denotes subtraction, $\oslash$ denotes division, and $\otimes$ denotes multiplication in the arithmetic $\boldsymbol{R}_{e10^{100}}$.

$$1000 \oplus 1000 = 2000$$
$$10^{90} \oplus 10^{90} = 2 \times 10^{90}$$
$$10^{200} \oplus 10^{10} = 10^{100}$$
$$1000 \otimes 1000 = 1000000$$
$$10^{90} \otimes 10^{90} = 10^{100}$$
$$10^{200} \otimes 10^{10} = 10^{100}$$
$$10^{90} \ominus 10^{80} = (10^{10} - 1)10^{80}$$
$$10^{200} \ominus 10^{10} = 10^{100}$$
$$(10^{200} + 1000) \ominus 10^{200} = 10^{100}$$
$$10^{200} \oslash 10^{10} = 10^{100}$$
$$10^{200} \oslash 10^{100} = 10^{100}$$

**Proposition 3.8.** For any natural number $L$, the arithmetic $\boldsymbol{A}_{eL}$ is a subarithmetics of the arithmetic $\boldsymbol{R}_{eL}$.

Although the arithmetic $\boldsymbol{R}_{eL}$ is much larger than the arithmetic $\boldsymbol{A}_{eL}$, they have many similar properties.

**Theorem 3.8.** For any natural number $L$, we have:
a) addition and multiplication are commutative in the arithmetic $\boldsymbol{R}_{eL}$;
b) addition in the arithmetic $\boldsymbol{R}_{eL}$ is not always associative;
c) multiplication in the arithmetic $\boldsymbol{R}_{eL}$ is always associative;
d) multiplication in the arithmetic $\boldsymbol{R}_{eL}$ is not always distributive with respect to addition.

*Proof* is similar to the proof of Theorem 3.1.

At the same time, the arithmetic $R_{eL}$ has properties such that the arithmetic $A_{el}$ does not have.

**Theorem 3.10.** A number $r$ from the arithmetic $R_{eL}$ has the inverse if and only if $|r| \geq 1/L$.

*Proof* is similar to the proof of Theorem 3.7.

Note that in the arithmetic $A_{el}$, numbers do not inverse numbers.

At the same time, division is defined for all numbers from the arithmetic $R_{eL}$ but 0.

Comparing arithmetics $R_{eL}$ and $R_L$, we see that, $R_{eL}$ has properties such that $R_L$ does not have.

For instance, if $m > n > L$ in the arithmetic $R_L$, then we have

$$m -_L n = h(g(m) - g(n)) = h(L - L) = h(0) = 0$$

In particular, it means that in the arithmetic $R_{10^{100}}$ considered above, we have

$$(10^{100} + 100) \ominus (10^{100} + 50) = 0$$

However, taking the arithmetic $R_{e10^{100}}$ considered above, we see that

$$(10^{100} + 100) \ominus (10^{100} + 50) = h((10^{100} + 100) - (10^{100} + 50)) =$$
$$h(100 - 50) = h(50) = 50$$

Another important difference between arithmetics $R_{eL}$ and $R_L$ emerges in division.

In the arithmetic $R_{e10^{100}}$, we have:

$$10^{200} \oslash 10^{10} = 10^{100}$$
$$10^{200} \oslash 10^{100} = 10^{100}$$
$$10^{120} \oslash 10^{110} = 10^{10}$$

At the same time, in the arithmetic $R_{10^{100}}$, we have

$$10^{200} \oslash 10^{10} = 10^{90}$$
$$10^{200} \oslash 10^{100} = 1$$
$$10^{120} \oslash 10^{110} = 1$$

Let us consider the subarithmetic $R_{eL}^+$ of all positive numbers from the arithmetic $R_{eL}$.

**Theorem 3.9.** For any natural number $L$, we have:

a) addition and multiplication are commutative in the arithmetic $R_{eL}^+$;

b) addition and multiplication are associative in the arithmetic $R_{eL}^+$;

c) multiplication in the arithmetic $R_{eL}^+$ is distributive with respect to addition;

d) addition and multiplication are monotone operations in the arithmetic $R_{eL}^+$.

*Proof* is similar to the proof of Theorem 3.2.

**Remark 3.4**. In the non-Diophantine arithmetics $R_L^+$, $R_L^+$, $R_{eL}$, and $R_{eL}^+$, it is possible to define such integral operations as sequential multiple addition, parallel multiple addition, sequential multiple multiplication and parallel multiple multiplication, as well as sequential and parallel exponents using analogous constructions in non-Diophantine arithmetics $A_L^+$, $A_L^+$, $A_{eL}$, and $A_{eL}^+$. These operations in $R_L^+$, $R_L^+$, $R_{eL}$, and $R_{eL}^+$ have properties similar to the proved properties in $A_L^+$, $A_L^+$, $A_{eL}$, and $A_{eL}^+$.

**Remark 3.5**. Using constructed non-Diophantine arithmetics, we can build different non-Grassmannian linear spaces for quantum mechanics [6]. In the discrete case, it might be resourceful using arithmetics $A_{eL}$ and $A_L$. In the continuous case, it might be resourceful using arithmetics $R_{eL}$ and $R_L$.

## 4. Perturbation theory based on non-Diophantine arithmetics

As noted in Section 2, the results for the second order of perturbation theory of QED depend on the divergent integrals given by formulas (4) and (5), where $I_1$ is the divergent contribution to the Feynman diagram describing the electron self energy, $I_2$ is the divergent contribution to the Feynman diagram describing the photon self energy, and $I_3$ is the divergent contribution to the Feynman diagram describing the electron-photon vertex.

While these integrals are not well defined, one can notice that the integrands in the integrals do not depend on hyperspherical angular variables. So, we can easily integrate over those variables. Instead of the Euclidean variables ($p_1$, $p_2$, $p_3$, $p_4$), we introduce the variable $p = (p_1^2 + p_2^2 + p_3^2 + p_4^2)^{1/2}$, which is the sum of $p_1$, $p_2$, $p_3$, and $p_4$ in the non-Diophantine arithmetic with the functional parameter $f(x) = x^2$ described in [11], and the hyperspherical angles ($\psi$, $\theta$, $\varphi$) where $\psi$ and $\theta$ have the range from 0 to $\pi$, while $\varphi$ has the range from 0 to $2\pi$ and define

$$p_4 = p\cdot\cos\psi;\ p_3 = p\cdot\sin\psi\cdot\cos\theta;\ p_1 = p\cdot\sin\psi\cdot\sin\theta\cdot\cos\varphi;\ p_2 = p\cdot\sin\psi\cdot\sin\theta\cdot\sin\varphi \quad (7)$$

Then $d^4p = dp_1 dp_2 dp_3 dp_4 = p^3 \sin^2\psi \sin\theta\, d\psi\, d\theta\, d\varphi\, dp$ and, after integration over ($\psi$, $\theta$, $\varphi$), we formally obtain $I_i = 2\pi^2 J_i$ ($i = 1, 2, 3$) where

$$J_1 = \int_0^\infty (p^2+l)^{-2} p^3 dp,\quad J_2 = \int_0^\infty (p^2+l)^{-2} p^5 dp,\quad J_3 = \int_0^\infty (p^2+l)^{-3} p^5 dp \quad (8)$$

In standard theory, those integrals are divergent, and in the literature, this is sometimes illustrated as follows. Let $J_i(p_{\max})$ ($i = 1, 2, 3$) be the integrals in formula (8) where the upper limit is not $\infty$ but $p_{\max}$. Then a simple integration gives that if $p_{\max}$ is very large, then we have

$$J_1(p_{\max})=\tfrac{1}{2}(\ln(p_{\max}^2/l)-1),\ J_2(p_{\max})=\tfrac{1}{2}(p_{\max}^2+2l\ln(p_{\max}^2/l)+l),\ J_3(p_{\max})=\tfrac{1}{2}(p_{\max}^2/l-3/2) \quad (9)$$

From the point of view of formal construction of QED, one should take the limits of those expressions when $p_{\max} \to \infty$ because, as noted in Secion 1, QED is based on conventional mathematics where there is no finite absolute maximum for the momentum. However, these limits do not exist. This is an indication that mathematically QED is not well defined. Then a question arises why, nevertheless, QED describes experimental data with a high accuracy.

The answer is as follows. Perturbation theory in QED starts from the bare electron mass $m_0$ and bare electron electric charge $e_0$. However, the description of experiment should involve not those quantities but real electron mass $m$ and real electron charge $e$. It has been proved that, in each order of perturbation theory, all singularities of unknown quantities $m_0$ and $e_0$ and all singularities of QED perturbation theory are fully absorbed by $m$ and $e$ such that the resulting formulas expressed in terms of $m$ and $e$ do not contain singularities anymore.

This property of QED is characterized such that QED is a renormalizable theory. As noted in Section 1, a very impressive property of QED is that it describes the electron and muon magnetic moments with the accuracy eight decimal digits. This result was achieved in the third order of perturbation theory, and so far, no comparable results in theory and experiment in higher orders have been obtained. It was also proved that electroweak theory and quantum chromodynamics also are renormalizable theories. At the same time, QED and those theories cannot answer the question whether the perturbation series in them are convergent or asymptotic. Also, the existing QFT versions of quantum gravity are not renormalizable.

Despite successes of renormalizable theories in describing experimental data, the above discussion shows that those theories are not well defined mathematically. As it is indicated in Section 1, one of the reasons is that they contain products of quantized fields at the same points. Also, different authors pointed out (see e.g., Chapter 24 in [8]) that, although from the formal point of view, the value of $p_{max}$ in QED should be infinitely large, it follows from physical considerations that this value is finite and plays an important role.

We now consider how the integrals in formulas (6) should be treated in non-Diophantine mathematics (NDM). A detailed description of NDM has been given in [6] and the basic facts of NDM have been described in Section 3. Let $S(x)$ be a set of integer, rational or real numbers $x$. Then, as noted in Section 3, in NDM there always exists a number $L$ with the following properties: if $x_1, x_2 \in S$ then the results of addition, subtraction and multiplication of $x_1$ and $x_2$ will be the same as in standard mathematics if $x_1$ and $x_2$ are much less than $L$ but can essentially differ from the results in standard mathematics is $x_1$ and/or $x_2$ are comparable to $L$.

Now let us consider how the integrals in formulas (6) can be treated in NDM. As noted above, the integrands in the integrals in formulas (3) and (4) do not depend on hyperspherical angular variables. Therefore, as explained at the beginning of this section, in standard mathematics one can immediately integrate over those variables and get formulas (8).

Since the modulus of each hyperspherical angular variable does not exceed $2\pi$ then the operations with those variables in standard mathematics and in NDM are the same. Therefore, in NDM it is also possible to integrate over the angular variables according to the same rules as in standard mathematics and get $I_i = 2\pi^2 J_i$ $(i=1, 2, 3)$. However, while in standard mathematics the quantities $J_i$ are not well defined because the integrals in formulas (8) diverge, we will now show that in NDM such integrals are well defined.

Consider, for example, the integral $J_1$. The Riemann sums for this integral are defined in the following way. We represent the interval $[0, \infty)$ as the union $[0, \infty) = \cup_0^\infty [p_i, p_{i+1})$ where $p_i = i\Delta p$ $(i= 0, 1, ...\infty)$ where $\Delta p > 0$. Then the Riemann sum for $J_1$ is

$$S(n) = \sum_{i=1}^{n} [(p_i^3)/(p_i^2 + l)^2]\Delta p \qquad (10)$$

and $J_1$ is the limit of $S(n)$ when $\Delta p \to 0$ and $n \to \infty$.

Let us note that $p$ and $l$ are the dimensionful quantities, and their dimensions depend on the used systems of units. For example, in SI the dimension of $p$ is $kg \cdot m/s$ while in the system of units $\hbar = c = 1$, which is often used in particle theory, the dimension is $1/length$. To obtain the corresponding descriptions in NDM, it is necessary to use non-Grassmannian linear spaces [6] where integer, rational and real numbers are dimensionless. For this reason we define $p=ax$ where $a$ is a constant having the dimension of momentum, and $x \in [0, \infty)$ is the dimensionless variable. Then we have

$$S(n) = \sum_{i=1}^{n} [(x_i^3)/(x_i^2 + b)^2]\Delta x \qquad (11)$$

where $\Delta x = \Delta p/a$, $p_i = ax_i$ and $b = l/a^2$.

Since all the terms in the sum (11) are positive, in standard theory $J_1$ diverges and $J_1$ is a limit of the sums $S(n)$ when $\Delta x \to 0$ and $n \to \infty$, then in standard theory, $\forall L > 0 \: \exists \delta > 0$ and $\exists n_0$ such that $S(n) > L$ for $\forall \Delta x < \delta$ and $\forall n > n_0$.

To eliminate the unwelcome divergence of the considered integrals, we use the non-Diophantine arithmetic $\boldsymbol{R}_L$ because in it, operations with numbers and functions cannot go beyond the boundary number $L$ in the positive direction and the boundary number $-L$ in the negative direction. At the same time, the basic results of contemporary physics in general and quantum theory in particular, which do not involve infinity in the form of divergence, are preserved in this new setting if we take $L$ sufficiently big because for all numbers in $\boldsymbol{R}_L$ from the interval $(L^{1/2}/2, -L^{1/2}/2)$, all basic arithmetical operations are the same as in the conventional Diophantine arithmetic $\boldsymbol{R}$ of real numbers. Note that if $L$ is very big, the interval $(L^{1/2}/2, -L^{1/2}/2)$ is sufficiently big.

Contemporary quantum physics is based on the Diophantine arithmetic because in it, all operations with numbers and functions are performed according to the rules of this arithmetic. Using operations from a non-Diophantine arithmetic in QFT, we obtain ND quantum physics.

When we utilize the non-Diophantine arithmetic $\boldsymbol{R}_L$ for building ND quantum physics, the Riemann sum (10) is transformed to the non-Diophantine Riemann sum (12).

$$S_L(n) = \sum_{i=1}^{n\oplus} [(p_i{}^{3\otimes}) \oslash (p_i{}^{2\otimes} \oplus l)^{2\otimes}] \otimes \Delta p \qquad (12)$$

Here $\oplus$ denotes addition, $\ominus$ denotes subtraction, $\oslash$ denotes division, $\Sigma^{\oplus}$ denotes multiple addition, and $\otimes$ denotes multiplication in the non-Diophantine arithmetic $\boldsymbol{R}_L$.

Taking the limit of $S_L(n)$ when $\Delta p \to 0$ and $n \to \infty$, we obtain a non-Newtonian integral [6].

$$J_{L1} = \int^{\otimes} [(p^{3\otimes}) \oslash (p^{2\otimes} \oplus l)^{2\otimes}] \otimes dp \qquad (13)$$

In ND quantum physics based on the non-Diophantine arithmetic $\boldsymbol{R}_L$, this integral is the counterpart of the integral $J_1$ that describes the electron self energy.

By the construction of the non-Diophantine arithmetic $\boldsymbol{R}_L$, the sum (12) cannot be larger than the number $L$. Consequently, the integral (13) also cannot be larger than the number $L$.

By the same techniques as before, we transform the Riemann sum (11) to the non-Diophantine Riemann sum (14) in the non-Diophantine arithmetic $\boldsymbol{R}_L$.

$$S_{LA}(n) = \sum_{i=1}^{n\oplus} [(x_i{}^{3\otimes})\oslash(x_i{}^{2\otimes} \oplus b)^{2\otimes}]\otimes\Delta x \qquad (14)$$

Here $\oplus$ denotes addition, $\ominus$ denotes subtraction, $\oslash$ denotes division, $\sum^{\oplus}$ denotes multiple addition, and $\otimes$ denotes multiplication in the non-Diophantine arithmetic $\boldsymbol{R}_L$.

Taking the limit of $S_L(n)$ when $\Delta x \to 0$ and $n \to \infty$, we obtain a non-Newtonian integral (Burgin and Czachor, 2020)

$$J_{LA} = \int^{\otimes} [(x^{3\otimes})\oslash(x^{2\otimes} \oplus b)^{2\otimes}]\otimes \mathrm{d}x \qquad (15)$$

In ND quantum physics based on the non-Diophantine arithmetic $\boldsymbol{R}_L$, this integral is the counterpart of the integral $J_1$ that describes the electron self energy,

By construction of the non-Diophantine arithmetic $\boldsymbol{R}_L$, the sum (14) cannot be larger than the number $L$. Consequently, the integral (15) also cannot be larger than the number $L$.

In a similar way, we can demonstrate that in the non-Diophantine arithmetic $\boldsymbol{R}_L$, for all integrals - $J_{L1}$ that describes the electron self energy, $J_{L2}$ that describes the photon self energy, and $J_{L3}$ that describes the electron-photon vertex, - we have the following inequalities

$$J_{L1} \leq L, \; J_{L2} \leq a^2 L, \; J_{L3} \leq L \qquad (16)$$

This shows that in contrast to standard mathematics where the values of the integrals $I_i$ ($i = 1; 2; 3$) are infinite, their counterparts $J_{L1}$, $J_{L2}$ and $J_{L3}$ in ND quantum physics based on the non-Diophantine arithmetic $\boldsymbol{R}_L$ are finite.

Note that elimination of infinities (convergence) in quantum physics can be based on the non-Diophantine arithmetic $\boldsymbol{R}_{eL}$ instead of $\boldsymbol{R}_L$ and utilize non-Grassmannian linear space over the non-Diophantine arithmetic $\boldsymbol{A}_L$ or $\boldsymbol{A}_{eL}$ instead of $\boldsymbol{R}_L$ or $\boldsymbol{R}_{eL}$.

## 5. Conclusion

In Section 2 we discussed examples of divergent integrals arising in the second order of perturbation theory in QED. Those examples demonstrate that, from the mathematical point of view, standard construction of QED is not well defined. Nevertheless, in many cases QED describes experimental data with a very high accuracy.

As explained in Section 4, the reason of such a situation is as follows. Perturbation theory in QED starts from the bare electron mass $m_0$ and bare electron electric charge $e_0$. However, the description of experiment should involve not those quantities but real electron mass $m$ and real electron charge $e$. It has been proved that, in each order or perturbation theory, all singularities of unknown quantities $m_0$ and $e_0$ and all singularities of QED perturbation theory are fully absorbed by $m$ and $e$ such that the resulting formulas expressed in terms of $m$ and $e$ do not contain singularities anymore.

This situation can be characterized such that the problems with mathematical justification have been swept under the carpet. An analogous situation takes place in other known renormalizable theories - electroweak theory and quantum chromodynamics.

As explained in Section 4, in the approach when QED is based on non-Diophantine mathematics (NDM), there are no divergent integrals. In this section, we discussed NDM analogs of integrals [6]. In standard approach to QED, the initial integrals are divergent but in the NDM approach their counterparts are well defined. It is necessary to keep in mind that another way to cope with infinities emerging in theoretical physics is not to eliminate them but to use the mathematical techniques allowing rigorously working with infinities (see e.g., [17]).

Therefore, in the approach when QED is based on NDM, the construction of the theory can be performed in full analogy with the standard construction, but the quantities $m$ and $e$ are now expressed not in terms of singular quantities but in terms of quantities which are well defined. In a similar way, it is possible to show that in the NDM approach, all other known renormalizable theories are mathematically well defined.